\documentclass[12pt]{article}
\usepackage[dvips]{graphics}
\usepackage{amssymb}

\newcommand{\nc}{\newcommand}
\nc{\figcap}[1]{\begin{quote}\refstepcounter{figure}
        {\bf Figure \thefigure}: {\small #1}\end{quote}}
\nc{\fig}[1]{\mbox{Fig.~\ref{#1}}}

\nc{\noi}{\noindent}

\nc{\bea}{\begin{eqnarray}}
\nc{\eea}{\end{eqnarray}}
\nc{\bean}{\begin{eqnarray*}}
\nc{\eean}{\end{eqnarray*}}
\nc{\ba}{\begin{array}}
\nc{\ea}{\end{array}}
\nc{\be}{\begin{equation}}
\nc{\ee}{\end{equation}}
\nc{\nn}{\nonumber}
\nc{\bra}[1]{\langle #1|}
\nc{\ket}[1]{|#1\rangle}
\nc{\av}[1] {\langle #1\rangle}
\nc{\vac}[1] {\langle 0| #1|0\rangle}
\nc{\amp}[2]{\langle #1|#2\rangle}
\nc{\da}{\dagger}
\nc{\pa}{\partial}
\nc{\ga}{\gamma}
\nc{\ep}{\epsilon}
\nc{\tf}{t_f}
\nc{\half}{\ensuremath{\frac{1}{2}}}
\nc{\hHH}{\hat H}
\nc{\ha}{\hat a}
\nc{\hO}{\hat O}
\nc{\hAA}{\hat A}
\nc{\hB}{\hat B}
\nc{\hG}{\hat G}
\nc{\hN}{\hat N}
\nc{\hU}{\hat U}
\nc{\hx}{\hat{x}}
\nc{\hp}{\hat{p}}
\nc{\hpsi}{\hat \psi}
\nc{\hphi}{\hat \phi}
\nc{\hpi}{\hat \pi}
\nc{\hpd}{\hat \psi ^\dagger}
\nc{\hE}{\hat E}
\nc{\hb}{\hat b}
\nc{\hc}{\hat c}
\nc{\hjo}{\hat j _0}
\nc{\hrho}{\hat \rho}
\nc{\leave}{\! \! \! \! \! / \, \,}
\nc{\intl}[1]{\int d\! #1 \,} 
\nc{\intll}[3]{\int _#1^#2 d\! #3 \,} 
\nc{\lm}{\lim _{y \rightarrow x}}

\nc{\scd}{\partial ^2 _{A_T}}
\nc{\fd}[1]{\frac{\delta }{\delta #1}} 
\nc{\pad}[1]{\frac{\partial}{\partial #1}} 
\nc{\refpa}[1]{(\ref{#1})} 

\nc{\calH}{\ensuremath{\mathcal{H}}}
\nc{\calD}{\ensuremath{\mathcal{D}}}
\nc{\calL}{\ensuremath{\mathcal{L}}}
\nc{\calO}{\ensuremath{\mathcal{O}}}
\nc{\hcalO}{\ensuremath{\hat \mathcal{O}}}
\nc{\calK}{\ensuremath{\mathcal{K}}}
\nc{\Tr}{\ensuremath{\mathrm{Tr}}}
\nc{\tr}{\ensuremath{\mathrm{tr}}}
\nc{\ra}{\rightarrow}
\nc{\lr}{\leftrightarrow}
\nc{\phistar}{\phi^*}
\nc{\etat}{\eta_T}
\nc{\het}{\hat E_T}
\nc{\hpt}{\hat \psi_T}
\nc{\hpdt}{\hat \psi ^\dagger_T}
\nc{\bart}{\bar{t}}
\nc{\barp}{\bar{p}}
\nc{\barT}{\bar{T}}
\nc{\hbarrho}{\hat{\bar{\rho}}}
\nc{\bga}{\ensuremath{\mbox{\boldmath{$\gamma$}}}}
\nc{\bsi}{\ensuremath{\mathbf{\sigma}}}
\nc{\bx}{\ensuremath{\mathbf{x}}}
\nc{\by}{\ensuremath{\mathbf{y}}}
\nc{\bz}{\ensuremath{\mathbf{z}}}
\nc{\bp}{\ensuremath{\mathbf{p}}}
\nc{\bn}{\ensuremath{\mathbf{n}}}
\nc{\bbp}{\ensuremath{\bar{\mathbf{p}}}}
\nc{\bP}{\ensuremath{\mathbf{P}}}
\nc{\hbA}{\hat{\ensuremath{\mathbf{A}}}}
\nc{\hbB}{\hat{\ensuremath{\mathbf{B}}}}
\nc{\bA}{\ensuremath{\mathbf{A}}}
\nc{\bJ}{\ensuremath{\mathbf{J}}}
\nc{\bB}{\ensuremath{\mathbf{B}}}
\nc{\bH}{\ensuremath{\mathbf{H}}}
\nc{\bM}{\ensuremath{\mathbf{M}}}
\nc{\bD}{\ensuremath{\mathbf{D}}}
\nc{\bE}{\ensuremath{\mathbf{E}}}
\nc{\hbE}{\hat{\ensuremath{\mathbf{E}}}}
\nc{\br}{\ensuremath{\mathbf{r}}}
\nc{\bj}{\ensuremath{\mathbf{j}}}
\nc{\bOm}{\ensuremath{\mathbf{\Om}}}
\nc{\om}{\omega}
\nc{\Om}{\Omega}
\nc{\sgn}{\mbox{sgn}}
\nc{\deltabar}{\mbox{$\delta\hspace*{-8pt}\vspace*{-8pt}-$}}
\nc{\gammat}{\tilde{\gamma}}
\nc{\binom}[2] {{#1\choose #2}}
\nc{\mub}{\bar{\mu}}
\nc{\rhob}{\bar{\rho}}
\nc{\Bb}{\bar{B}}
\nc{\Jb}{\bar{J}}
\nc{\Mb}{\bar{M}}
\nc{\Tb}{\bar{T}}
\nc{\sbar}{\bar{s}}
\nc{\betab}{\bar{\beta}}
\nc{\hj}{\hat j}
\nc{\hQ}{\hat Q}
\nc{\hJ}{\hat J}
\nc{\hA}{\hat A}
\nc{\hH}{\hat H}
\nc{\de}{\delta}
\nc{\leri}{\leftrightarrow}
\nc{\llabel}[1]{\label{#1}\marginpar{#1}}

\nc{\bc}{\begin{center}}
\nc{\ec}{\end{center}}
\nc{\inv}[1]{\frac{1}{#1}}

\newlength{\overeqskip}
\newlength{\undereqskip}
\nc{\eq}[1]{\mbox{Eq.~(\ref{#1})}}
\nc{\eps}{\epsilon}
\nc{\goto}{\rightarrow}

\nc{\cF}{{\cal F}}
\nc{\cG}{{\cal G}}
\nc{\cH}{{\cal H}}

\begin{document}
%
%
%
%
%
%
\thispagestyle{empty} 
%
%
%
%
%
%
%
%
%
\vspace{-5mm} 
\begin{center} 
\baselineskip 1.2cm 
{\Huge\bf  On the Preparation of Pure States in Resonant Microcavities
}\\[5mm] 
\normalsize 
\end{center} 
{\centering 
{\large Per K. Rekdal\footnote{Email address: p.k.rekdal@ic.ac.uk.}$^{,a}$, 
{\large Bo-Sture K. Skagerstam\footnote{Email 
address: boskag@phys.ntnu.no.}$^{,b,c}$ 
and 
{\large Peter L. Knight\footnote{Email address: p.knight@ic.ac.uk.}$^{,a}$} 
 \\[5mm] 
$^{a}~$QOLS, Blackett Laboratory, Imperial College, London SW7 2BW, United Kingdom\\[1mm]
$^{b}~$Department of Physics, 
The Norwegian University of Science and Technology,   
N-7491 Trondheim, Norway\\[1mm]
$^{c}~$Microtechnology Center at Chalmers MC2, Department of
  Microelectronics 
and Nanoscience, Chalmers University of Technology and 
G\"{o}teborg University, S-412 96, G\"{o}teborg, Sweden \\[5mm]
} } }
%
%
%
%
\begin{abstract} 
\normalsize 
%
%

\noindent We consider the time evolution of the radiation field $(R)$ and a two-level atom $(A)$
in a resonant microcavity in terms of the Jaynes-Cummings model with  an initial general pure
 quantum state for the radiation field. It is then shown, using the Cauchy-Schwarz inequality and also a Poisson
resummation technique,
     that {\it perfect} coherence of the atom can in general never be achieved.
     The atom and the radiation field are, however,
     to a good approximation in a pure state $|\psi \rangle_{A\otimes R}=|\psi \rangle_A\otimes|\psi \rangle_R$ in the middle of what has been 
	traditionally called the
     ``collapse region'', independent of the initial state of the atoms, provided that the 
initial pure state of the radiation field has a photon number probability  distribution which 
is  sufficiently peaked and phase differences that do not vary
significantly around this peak.
     An approximative analytic expression for the quantity 
	$\Tr[ \, \rho^2_{A}(t) \, ]$,
     where $\rho_{A}(t)$ is the reduced density matrix for the atom, is derived. We also show that under quite general circumstances an initial entangled  pure state will be disentangled to the pure state $|\psi \rangle_{A\otimes R}$.

\end{abstract} 

\newpage

The Jaynes-Cummings (JC) model \cite{Jaynes63} is one of the simplest but non-trivial examples of two interacting quantum systems. It is an important fundamental theoretical model of the interaction between a two-level atom and a second-quantized single-mode electromagnetic field (for a review see e.g. Ref.\cite{shore93}). 
The model is exactly solvable and realizable by experiments e.g. involving the passage of single atoms through a superconducting microwave cavity (see e.g. Refs.\cite{Walter88}). A direct experimental verification of the one-mode field quantization in  a microcavity, as assumed in the JC model, has actually been carried out \cite{BSMDHRH_96}. Experiments on ion-traps \cite{meekhof&96} provide for an alternative arena in which case JC-like models again are important theoretical tools in describing the relevant physics. Recent developments of nano-electronic devices has also led to an experimental realization of circuits with superconducting Josephson junctions which behaves as two-level systems \cite{nakamura&99}. Entangling such a device to a resonator naturally leads to a description in terms of an effective JC model (see e.g. Refs.\cite{armour&99}). The results of the present paper to be discussed below may therefore have a quite broad range of potential applications.

    In spite of its apparent simplicity, the JC model has led to many non-trivial and unexpected results through the years as e.g. the well-known phenomena of collapses and revivals of the atomic population inversion \cite{Meystre75}. In Ref. \cite{Gea90} it was found that the atom is to a good approximation in a pure state $|\psi \rangle_{A\otimes R}$ in 
the middle of the "collapse region" provided the initial state of the radiation field is 
a coherent state. It was shown \cite{phoenix91} that the existence of this particular pure state can easily be demonstrated
in terms of the properties of the von Neumann entropy in quantum mechanics. It has also been 
shown that the appearance of this pure atomic state is independent of the form of the initial pure 
atomic state \cite{banac90}. For mixed states of the radiation field it is also possible to transfer coherence from atomic states to the state of 
the radiation field \cite{moya95}.
Here we will show that the purification of the atomic state is actually independent of the
nature of the initial {\it pure} state of the radiation field provided that the corresponding
number-operator probability distribution is sufficiently peaked and that phase differences of this state
are slowly varying around this peak. We will, however, at this moment restrict ourselves to initial states
which are not entangled and give some comments on entangled initial states in the end of this paper.

     The electromagnetic interaction between a two-level atom, with level  
     separation $\omega_0$, and a single mode  of the  
     radiation field in a cavity with frequency $\omega$ is described, in the rotating wave  
     approximation, by the JC Hamiltonian  \cite{Jaynes63}  (where we take $\hbar =1$)

\begin{equation}  \label{eq:JCH}  
   H=\omega a^{\da} a+\frac12\omega_0\sigma_z+g(a\sigma_++a^{\da}\sigma_-)~~,  
\end{equation}

     \noi 
     where the coupling constant $g$ is proportional to the dipole  
     matrix element of the atomic transition.  
     Here we make use of the Pauli matrices to describe the two-level atom and the  
     notation $\sigma_\pm=(\sigma_x\pm i\sigma_y)/2$.  The second-quantized  
     single mode electromagnetic field is described in a conventional  
     manner by means of an annihilation (creation) operator $a$ ($a^{\da}$),  
     where we have suppressed the cavity mode labels. For $g=0$ the  
     atom-field states $|n,s \rangle =|n \rangle \otimes |s \rangle$ are characterized by  
     the quantum numbers $n=0,1,\ldots$ of the oscillator and $s=\pm$
     for the atomic levels with energies $E_{n,\pm}= \omega n \pm \omega_0 /2$. 

For reasons of simplicity
we will now consider a resonant system, i.e. $\omega = \omega_0$.
We also assume that initially the atom is in the excited state $|+ \rangle$ and the 
      radiation field is in a general pure 
      state $|\gamma \rangle = \sum_{n=0}^{\infty} \sqrt{p_n} \, 
      e^{i \alpha_n} |n \rangle$. Hence, the initial state of the system is 
      $|\psi(0) \rangle = |+ \rangle \otimes |\gamma \rangle$. 
	The solution to the Schr\"{o}dinger 
      equation is then 
\be
 \label{psi_t}
  |\psi(t) \rangle 
               = \sum_{n=0}^{\infty} \sqrt{ p_n } \, e^{i \alpha_n} \, e^{-i\omega(n+1/2)t} 
                     \left ( ~ \cos( gt \sqrt{n+1} ) |n,+\rangle  - i 
                                \sin( gt \sqrt{n+1} ) |n+1,-\rangle ~ \right ) ~~.
\ee
\noi
      In order to describe the evolution of the atom alone it is convenient to
      introduce the reduced density matrix $\rho_{A}(t) = \Tr_{\gamma} [ \; |\psi(t) \rangle 
      \langle \psi(t)| \; ]$, where the trace is over a complete set of radiation field states.
One easily finds that 
\be
\rho_{A}(t)= p_{+}(t)|+ \rangle 
      \langle +|+ p_{-}(t)|- \rangle 
      \langle -|+c(t)|+\rangle 
      \langle -|+ c(t)^{*}|-\rangle 
      \langle +|~~.
\ee
\noindent Here $p_{+}(t)=1-p_{-}(t)=\sum_{n=0}^{\infty} ~ p_n \cos^2  ( gt \sqrt{n+1} )$ is the well known form for the revival probability and
\be
\label{defofc}
c(t) \equiv -i\sum_{n=0}^{\infty}\sqrt{p_{n} p_{n-1}}\cos(gt\sqrt{n+1})
\sin(gt\sqrt{n})e^{i(\alpha_{n} -\alpha_{n-1})}e^{-i\omega t}~~.
\ee

\noindent
 By straightforward algebra one can now verify that 
\be
\label{rhosquared}
\Tr [ \; \rho_{\gamma}^2(t) \; ] = \Tr [ \; \rho_{A}^2(t) \; ]= p_{+}^2(t)+p_{-}^2(t) +2|c(t)|^2 ~~~,
\ee
where the radiation field density matrix is $\rho_{\gamma}(t) = \Tr_{A} [ \; |\psi(t) \rangle \langle \psi(t)| \; ]$, and where the trace is over the atomic states. This means that purity
occurs in both subsystems at the same time and precisely at the same rate. We notice that, in general, the $\omega$ dependence cancels out in $\Tr [ \; \rho_{\gamma}^2(t) \; ]$.
For a coherent state $|z\rangle$, with $z=|z|\exp(i\alpha)$, we now have that $\alpha_n=n\alpha $ with  photon-numbers  that are Poisson distributed, i.e. $p_n = \exp(-|z|^2)|z|^{2n}/n!$. The $\alpha$ dependence
    will therefore also cancel out in $\Tr [ \; \rho_{\gamma}^2(t) \; ]$ for any initial state $|\gamma \rangle$ which has the same phase dependence as a coherent state.
 By a tedious but straightforward calculation, the identity $\Tr [ \; \rho_{\gamma}^2(t) \; ] = \Tr [ \; \rho_{A}^2(t) \; ]$ can actually be shown to be valid for any initial pure state of the form $|\psi(0) \rangle = (a|+ \rangle + b|- \rangle)\otimes |\gamma \rangle $ with $|a|^2 + |b|^2 =1$.

\vspace{25mm}
\begin{figure}[htp]  
\unitlength=1mm
\begin{picture}(110,80)(0,0)
\includegraphics{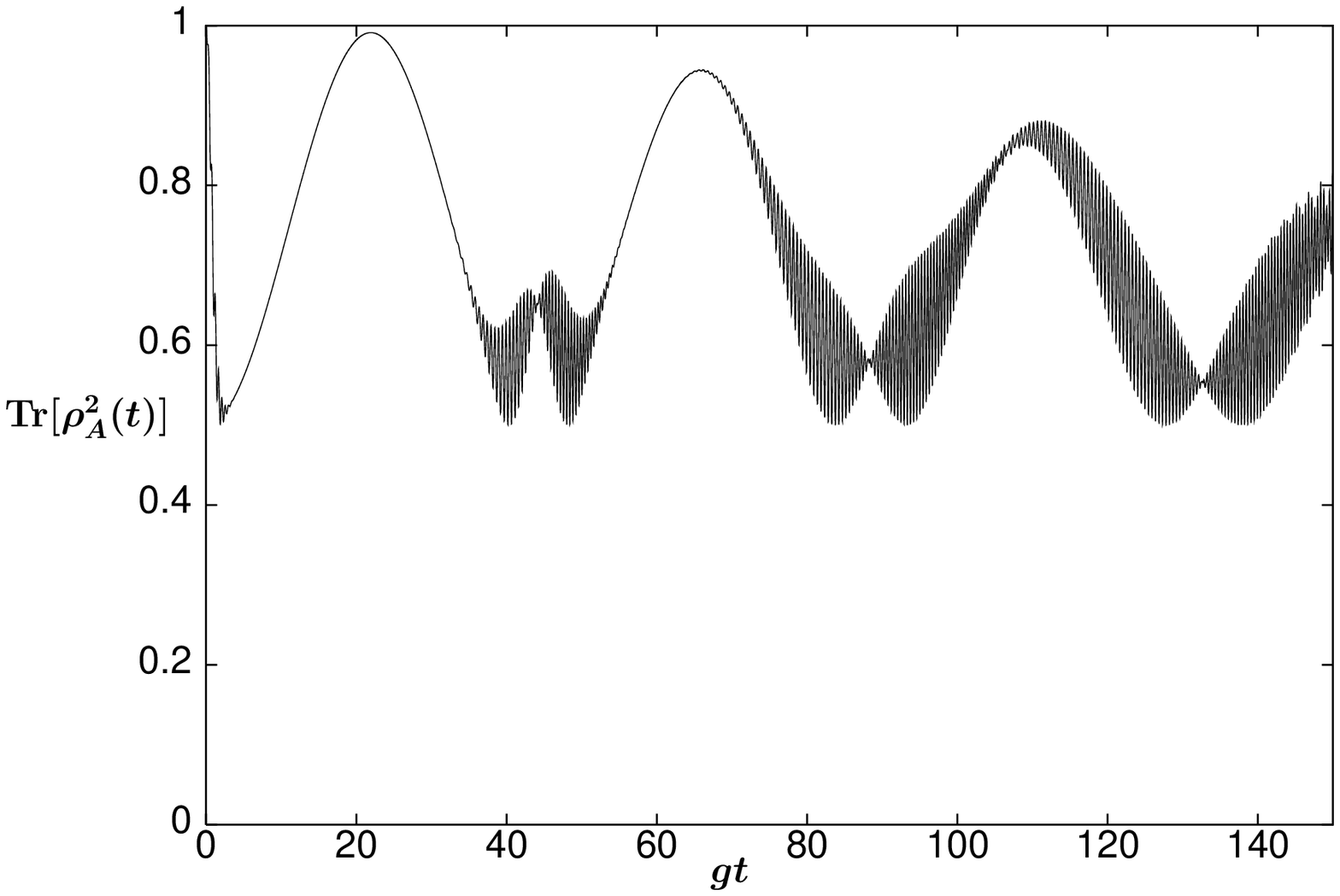}
\end{picture}
\caption[]{\protect\small  
   This figure shows $\Tr [ \, \rho_{A}^2(t) \, ]\; (= \Tr [ \, \rho_{\gamma}^2(t) \, ] )$ as a function of $gt$  with an initial coherent state of the radiation field with $|\alpha|^2=49$ based on the exact numerical results due to \eq{rhosquared}. This equation with a Gaussian photon-number distribution $p_n$, with mean ${\bar n}=|\alpha|^2$ and standard deviation $\sigma_n= \sqrt{{\bar n}}$,  describes the data  with a very high numerical accuracy. With our choice of parameters and within the accuracy of the figure shown, \eq{rhosquared} and the approximative expression of  $\Tr [ \, \rho_{A}^2(t) \, ]$ 
according to Eqs.(\ref{tr_2_peaked_analytisk}) and (\ref{w_nu_poisson}) agree.  As the mean
number of photons gets larger $\Tr [ \, \rho_{A}^2(t\simeq t_0 =\pi\sqrt{{\bar n}}) \, ]$ approaches one at an exponential rate
as can be seen from \eq{peak}. 
}
\label{tr_2_sml_FIG}
\end{figure}

      The purity of the atomic state can then be  determined considering 
the time evolution of the quantity $\Tr [ ~ \rho_{A}^2(t) ~ ] $\cite{Gea90}. 
A necessary and sufficient condition for the ensemble to be described in terms of a pure state 
is then that $\Tr [ ~ \rho_{A}^2(t) ~ ] = 1$, in which case clearly a state-vector description of each individual system of the ensemble is possible. On the other hand, for a two-level system, a maximally mixed ensemble corresponds to $\Tr [ ~ \rho_{A}^2(t) ~ ] = 1/2$.
Alternatively one may consider the von
Neumann entropy of the density matrix  $\rho_{A}(t)$ \cite{phoenix91} or simply its 
eigenvalues \cite{banac90}. One can also make use of the Schmidt decomposition of composite system (see e.g. Ref.\cite{eckert95}) to reach the same results as given below. 

The definition of $c(t)$ according to \eq{defofc}
 above now suggests an immediate interpretation in terms of an $l_2$ Hilbert space scalar product
of two complex vectors $a$ and $b$ with components  $a_n = \sqrt{p_n}\cos(gt\sqrt{n+1})\exp(i\alpha_{n})$ and  $b_n =\sqrt{p_{n-1}}\sin(gt\sqrt{n})\exp(i\alpha_{n-1})$ respectively with $b_0=0$. The Cauchy-Schwarz inequality then tells us that $|c(t)|^2\leq p_{+}(t) p_{-}(t)$ with equality if and only if the vectors $a$ and $b$ are parallel, i.e. $a_n = \beta (t)  b_n$ where $|\beta (t)|^2=p_{+}(t)/p_{-}(t)$. 
In general we then see that $1/2 \leq \Tr [ ~ \rho_{A}^2(t) ~ ] \leq 1$. The condition 
$a_n = \beta (t) b_n$ can be used to find a particular $p_n$ such that $\Tr [ ~ \rho_{A}^2(t) ~ ]=1$ at a fixed time $ t=t_f $. In order to see this let us assume that $\alpha_n = \alpha n + \alpha_0$. At $ t=t_f $ we then obtain a recursion formula $p_n \cos^2(gt_f\sqrt{n+1})=|\beta (t_f)|^2p_{n-1}\sin^2(gt_f\sqrt{n})$.  Depending on the parameter $gt_f$ this recursion formula can be solved. The corresponding probability distribution can now be used as an initial distribution and is  then such that $\Tr [ ~ \rho_{A}^2(t=t_f) ~ ]=1$ by construction. The time parameter $t_f$ is, however, not in general related to any natural revival time of the system.
%
%

In order to proceed in a more general setting, we  assume that the distribution $p_n$, with mean ${\bar n}$ and variance $\sigma_n$, is peaked around $n=\bar{n}$ or more precisely   $\sigma_n/\bar{n}\ll 1$. We observe that $\sigma_n$ should, however, not be arbitrarily small in order to have a non-zero $c(t)$. An explicit condition on $\sigma_n$ will be discussed below using Poisson resummation techniques.
Following the analysis of Ref.\cite{Gea90}, and by considering time-scales  $t$ close to half of the first revival time $t_{rev}=2\pi \sqrt{\bar{n}+1}/g$, i.e. $t\simeq t_0 \equiv t_{rev}/2$,  one finds that
$\Tr [ ~ \rho_{A}^2(t) ~ ]= 1/2 + 2|c(t)|^2$ using that $p_{\pm}(t\simeq t_0)=1/2$.
As long as $g|t-t_0 |<< 2\sqrt{\bar{n}+1}$, we can make use of the fact that $gt\sqrt{n+1} = gt\sqrt{n}+ \pi/2$ \cite{Gea90} and we find that $|c(t)|^2 =1/4$ provided $\sqrt{p_n}\exp(i\alpha_{n}) = \beta \sqrt{p_{n-1}}\exp(i\alpha_{n-1})$ for all $n$ with a complex phase $\beta $ ($|\beta | =1$). This is possible with high accuracy only if $ i)$ $p_n$ is non-zero (and constant) for a finite range of $n$ as we actually have assumed, and  $ ii)$ that the phases $\alpha _n$ are of the form $\alpha _n =\alpha n + \alpha _0$, i.e. the phase differences $\alpha _n -\alpha _{n-1}$ do not vary
significantly around $n=\bar{n}$. Under such circumstances we then see that $\Tr [ ~ \rho_{A}^2(t) ~ ]=1$ and the atom is in a pure state at $t\simeq t_0$.

\begin{figure}[htp]  
\unitlength=1mm
\begin{picture}(110,80)(0,0)
\includegraphics{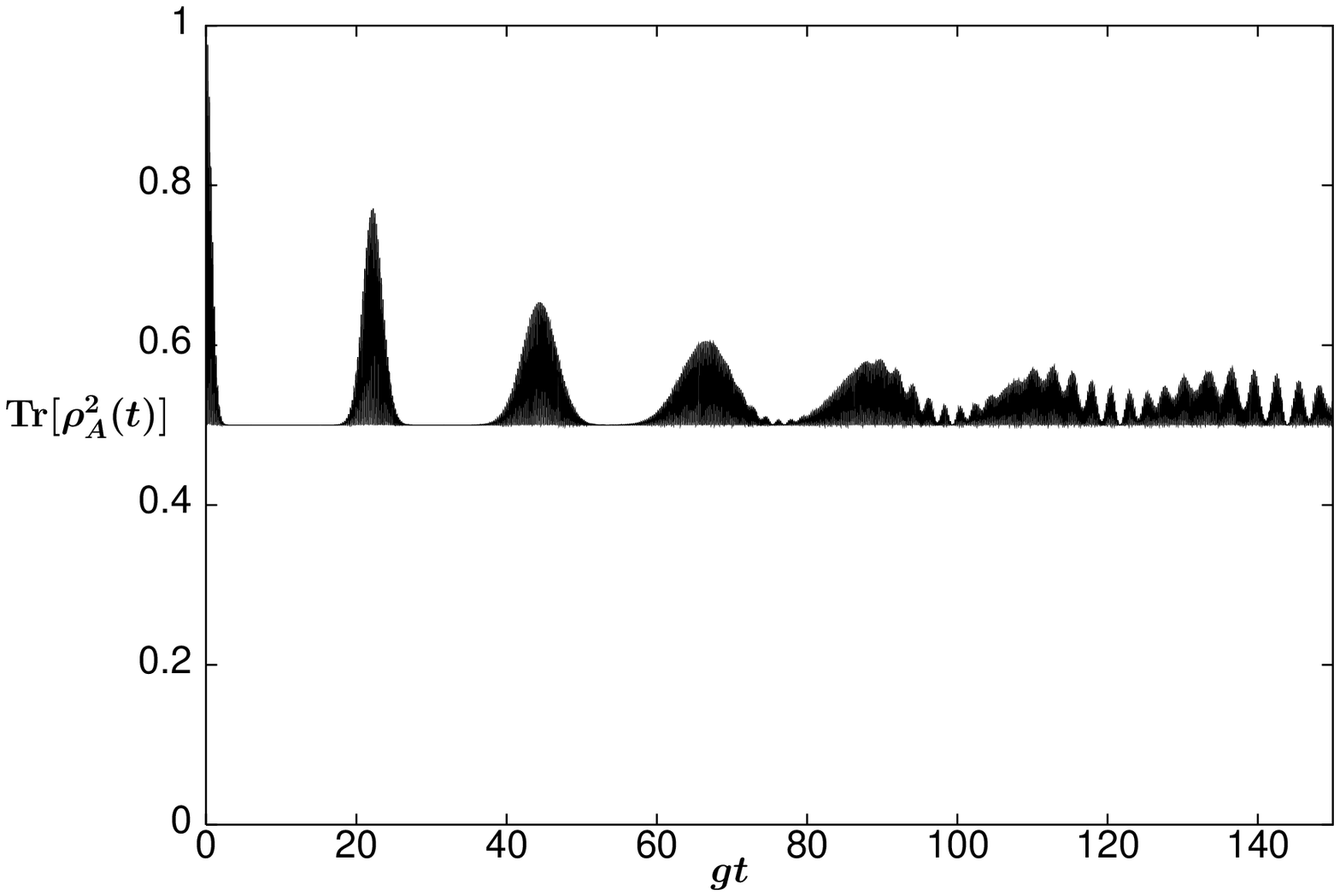}
\end{picture}
\caption[]{\protect\small  
    The quantity $\Tr [ \, \rho_{A}^2(t) \, ]\; (= \Tr [ \, \rho_{\gamma}^2(t) \, ] )$ 
           as a function of $gt$  when the initial state of the radiation field is  a Schr\"odinger cat state
           where $p_n = |\alpha|^{2n} \left [ 1 + (-1)^n \cos \phi \right ]/[ \, 
           n!  ( e^{ |\alpha|^2} +  e^{-|\alpha|^2} \cos \phi )  \, ]$ with 
           $|\alpha|^2=49$ and $\phi=0$. Purity of the atom occurs in this case for $gt=0$ only.
}
\label{tr_2_cat_FIG}
\end{figure}

In Fig.\ref{tr_2_sml_FIG} we illustrate the behaviour of $\Tr [ \, \rho_{A}^2(t) \, ]$ in the case of an initial coherent state for the radiation field with ${\bar n}\gg 1$. The distribution $p_n$ can the be approximated with a sharp Gaussian distribution obeying the condition $ i)$ above with high accuracy.  In Fig. \ref{tr_2_cat_FIG} we illustrate the behaviour of  $\Tr [ \, \rho_{A}^2(t) \, ]$ in the case of a pure state with $p_n$
which varies rapidly  in terms of a Schr\"{o}dinger cat state here chosen to be of the form $|\gamma \rangle =N(|\alpha \rangle+\exp(i\phi) |-\alpha \rangle)$ with $|N|^2=1/(2+2\cos(\phi)\exp(-2|\alpha|^2)$ (see e.g. Ref.\cite{gerry97}). For such a state  it follows from the analysis above that $|c(t)|$ cannot reach its maximal value $1/2$ as is also clearly exhibited in Fig. \ref{tr_2_cat_FIG}. The behaviour of  Schr\"{o}dinger cat states at $t= t_0$ including the effects of damping has been studied in more detail in Ref.\cite{skager&2001}.

It is now clear from the definition \eq{defofc} of $c(t)$  
that if the phases $\alpha_n$ are rapidly varying functions of $n$, $c(t)$ will be small or zero. If in particular the phases are random the state of the radiation field will correspond to a mixture instead of a pure state  and we have to average over the phases of $\rho_A(t)$ before computing its square.  We then obtain $\Tr [ \, \rho_{A}^2(t) \, ]= 1+2p_{+}(t)(p_{+}(t)-1)$
with purity of the atomic state only at times $t$ such that $p_{+}(t)=1$ or $0$.
In such a case we also notice that
$\Tr [ \, \rho_{A}^2(t) \, ]$ in general is different from $\Tr [ \, \rho_{\gamma}^2(t) \, ]$. The purity of the state of the radiation field is therefore required in order to purify a general initial mixture of the atom. For related studies of purification in the JC model see Refs.\cite{knight&etc&2001}.

    The quantity $\Tr [ \, \rho_{A}^2(t) \, ]$ can now be re-written in a form where 
    quantum revivals become explicitly by making use of a Poisson summation technique \cite{Schleich93}.  As above, we  consider a probability distribution $p_n$ is peaked around $n=\bar{n} $ with $\sigma_n/\bar{n}\ll 1$. If the phase differences $\alpha_n-\alpha_{n-1}$ are slowly varying functions of $n$ around $n=\bar{n}\gg 1$, i.e. we assume that $\alpha_n  \simeq \alpha n + \alpha_0$ for $n\simeq {\bar n}$, we then see
that $|c(t)|$ according to \eq{defofc} can be written in the form

\be
\label{absc(t)}
|c(t)|= \sum_{n=0}^{\infty}\sqrt{p_{n} p_{n-1}}\cos(gt\sqrt{n+1})\sin(gt\sqrt{n}) \approx \frac{1}{2}(S_s(t)-S(t))~~~,
\ee

%
 

%
%
%
%

 \noindent and we can then write $\Tr[ \, \rho_{A}^2(t) \, ]$ in the following convenient form

\be
\label{startingpoint}
\Tr[ \, \rho_{A}^2(t) \, ]\approx \frac{1}{2}(1+ S^2_c(t) + S^2_s(t)+ S^2(t) - 2S(t)S_s(t))~~~,
\ee

  \noi where we have defined the functions

\be
\label{defs1}
S_c(t) = \sum_{n=0}^{\infty}p_n\cos(2gt\sqrt{n+1})~~,~~S_s(t) = \sum_{n=0}^{\infty}p_n\sin(2gt\sqrt{n+1})~~~,
\ee
and
\be
\label{defs2}
S(t) = \sum_{n=0}^{\infty}p_n\sin(gt/2\sqrt{n+1})~~~.
\ee

\noindent  An exact Poisson resummation technique \cite{Schleich93} now enables us to e.g. write 

\be
S_c(t) + iS_s(t) = \sum_{\nu=-\infty}^{\infty} f_{\nu}(t) + p_0\frac{e^{2igt}}{2}~~~,
\ee

\vspace{10mm}

\noindent where

\be
\label{defoff}
f_{\nu}(t) = \int_{0}^{\infty}dn\;p(n)e^{2iS_{\nu}(n)} ~~~.
\ee
Here $p(n)$ is an analytical continuation of $p_n$ and $S_{\nu}(n)=\pi\nu n - gt\sqrt{n+1}$. If $p(n)$ varies slowly as compared to the variation of $S_{\nu}(n)$ one can evaluate $f_{\nu}(t)$ for $\nu \neq 0$ by a stationary phase approximation \cite{Schleich93}. For a Gaussian distribution  with mean ${\bar n}$ and variance $\sigma_n^2$, i.e. $p(n)=\exp(-(n-{\bar n})^2/2\sigma_n^2)/\sqrt{2\pi\sigma_n^2}$, this corresponds to the condition $\sigma_n^2\gg ({\bar n}+1)/\pi\nu$, provided ${\bar n}\gg 1$ . The contribution from $\nu = 0$ can be estimated by again considering a Gaussian distribution with ${\bar n} \gg 1$ in which case one finds that 

\be
\label{approx1}
\int_{0}^{\infty}dn\;p(n) e^{2igt\sqrt{n+1}}\approx e^{ -( gt\sigma_n)^2/2{\bar n}}e^{2igt\sqrt{{\bar n}+1}} ~~~.
\ee
The stationary phase condition on $S_{\nu }(n)$ now immediately leads to the general revival times $t_{rev}=2\pi\nu\sqrt{{\bar n}+1}/g$. Applying the same techniques to $S(t)$ according to \eq{defs2} leads to revival times which are of the order ${\bar n}$ larger than the those of $S_c(t)$ and $S_s(t)$ and we therefore make the approximation
\vspace{15mm}
\begin{figure}[htp]  
\unitlength=1mm
\begin{picture}(110,80)(0,0)
\includegraphics{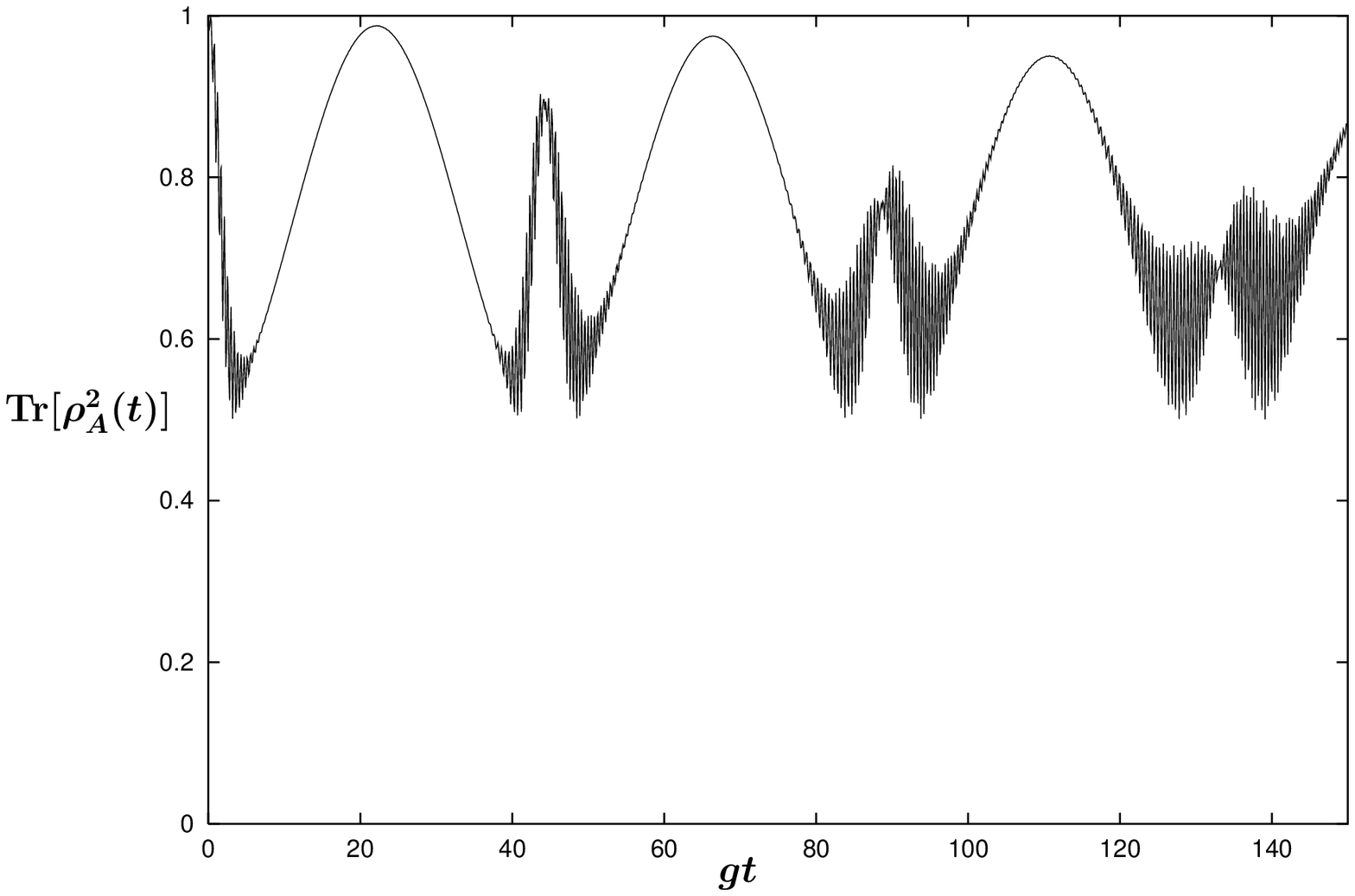}
\end{picture}
\vspace{-5mm}
\caption[]{\protect\small  
   This figure shows illustrates the effect of squeezing on $\Tr [ \, \rho_{A}^2(t) \, ] ~ 
    (=\Tr [ \, \rho_{\gamma}^2(t) \, ])$ as a function of $gt$  when the initial state of the radiation field is  a squeezed coherent state $|\gamma \rangle =S(r)|\alpha \rangle$ with $r=0.75$ and a real $\alpha~(\approx 14.72)$ chosen such that the mean number of photons ${\bar n}=49$ is the same as in \fig{tr_2_sml_FIG}. The photon-number distribution is well approximated by a "squeezed" Gaussian distribution with mean ${\bar n}$ and $\sigma_n^2= {\bar n}^{0.65}$. }
\label{tr_2_gaussian_FIG}
\end{figure}

\be
\label{approx2}
\sum_{n=0}^{\infty}p_n\sin \bigg ( \frac{gt}{2 \sqrt{n+1}  } \bigg )\approx \int_{0}^{\infty}dn \;p(n) \sin \bigg ( \frac{gt}{2 \sqrt{n+1}  } \bigg )\approx e^{ - ( gt\sigma_n)^2/32{\bar n}^{3} }\sin \bigg ( \frac{gt}{2 \sqrt{{\bar n+1}}  } \bigg ) ~~~,
\ee
 where we again have made use of Gaussian distribution and ${\bar n}\gg 1$. The approximations Eqs.(\ref{absc(t)}), (\ref{approx1}) and (\ref{approx2}) together with the stationary approximation above enables us to write the quantity $\Tr [ \;  \rho_{A}^2(t) \; ]$ in an analytical form, i.e.
\be \label{tr_2_peaked_analytisk} 
   \Tr [ ~ \rho_{A}^2(t) ~ ] \approx \frac{1}{2} + \frac{1}{2}  ~ 
                             e^{ - ( gt\sigma_n)^2/16{\bar n}^{3} }
      \sin^2 \bigg ( \frac{gt}{2\sqrt{{\bar n}+1}} \bigg ) 
                           + \sum_{\nu=0}^{\infty} w_{\nu}(t) ~,
\ee

\noi where 
\be \label{w_0_felles}
  w_{0}(t) = \frac{1}{2} \left [ ~ e^{ -(gt\sigma_n )^2/{\bar n} }  - 2  \, e^{ - (gt\sigma_n)^2(1+1/16{\bar n}^2)/2{\bar n} } \sin \bigg ( \frac{gt}{2 \sqrt{{\bar n}+1}} \bigg ) 
                                                      \sin ( 2 gt \sqrt{{\bar n}+1} ) \right ] ~,
\ee
 \noi 
 and for $\nu \geq 1$ 
\bea \label{w_nu_peak}
      w_{\nu}(t) &=& \frac{1}{2} \bigg [ ~ \, \bigg ( \frac{gt}{\pi \, \sqrt{2\nu^3}} \bigg )^2 ~ \,
                                      p^2 \bigg ( \, ( \frac{gt}{2 \pi \nu} \, )^2 \bigg )   \\ \nonumber  
                                       &-& 2 \, \frac{gt}{\pi \, \sqrt{ 2\nu^3} } ~
                                       p \bigg ( \, ( \frac{gt}{2 \pi \nu} \, )^2 \bigg )  
                                       \; e^{ - ( gt\sigma_n)^2/32{\bar n}^{3} } 
                                  \sin \bigg ( \frac{gt}{2 \sqrt{{\bar n}+1}  } \bigg ) 
                                  \sin \bigg ( \frac{(gt)^2}{2 \pi \nu} - \frac{\pi}{4} \bigg ) ~ 
                        \bigg ] ~.
\eea
 \noi
      Apart from the approximations used in order to obtain Eqs.(\ref{approx1})-(\ref{approx2}),  the Poisson resummed expression  for $\Tr [ \;  \rho_{A}^2(t) \; ]$  as given by \eq{tr_2_peaked_analytisk}  with  Eqs.(\ref{w_0_felles})-(\ref{w_nu_peak}) is valid for any distribution $p(n)$ provided that ${\bar n}\gg 1 $ and 
\be
\label{condition}
({\bar n}+1)/\nu\pi\ll \sigma_n^2 \ll {\bar n}^2~~~.  
\ee
With this condition, which is always satisfied for a Poissonian distribution,  we have also verified
 through  numerical calculations that
the various approximations leading to \eq{tr_2_peaked_analytisk}  describe a much larger class of $p_n$-distributions than just Gaussian ones. For ${\bar n}\gg 1$ we can now extract the leading behaviour of $  \Tr [ ~ \rho_{A}^2(t) ~ ] $ close to $t=t_{rev}/2$, i.e.
\be
\label{peak}
\Tr [ ~ \rho_{A}^2(t\simeq \frac{1}{2}t_{rev})  ] \simeq \frac{1}{2}\bigg( 1+
e^{ - \sigma_n^2 \pi^2/16{\bar n}^2 } + e^{ - \sigma_n^2 \pi^2 }-2e^{- \sigma_n^2 \pi^2 /2}\sin(2\pi{\bar n}) \bigg)~~~,
\ee
which show the exponential approach to purity.
     \noindent These results shows that the quantity $\Tr [ \;  \rho_{A}^2(t) \; ]$ exhibits revivals
     not only for an initial coherent state, as one is lead to believe from Ref. \cite{Gea90}, 
     but for any initial state  of the radiation field in the form of a pure  state
     $|\gamma \rangle = \sum_{n=0}^{\infty} \sqrt{p_n} \, e^{i \alpha_n} |n \rangle$ 
     with a sufficiently peaked probability distribution $p_n$, satisfying \eq{condition}, and with phase differences $\alpha_n - \alpha_{n-1}$ varying slowly around this peak.

     In particular, if the radiation field is initially in a coherent state with average photon number 
     ${\bar n} = |\alpha|^2$ as in Ref. \cite{Gea90}, then $p_n$ is Poisson distributed 
     and reduces to the Gaussian distribution with variance $\sigma_n^2={\bar n}$ if ${\bar n}\gg 1$.  In this case, \eq{w_nu_peak} can be reduced to
\bea \label{w_nu_poisson}
  w_{\nu}(t) &=& \frac{1}{2} \bigg [ ~ \frac{(gt)^2}{4\pi ^3 \nu ^3 {\bar n}} \; e^{ -\frac{(gt- g{\bar t}_{\nu})^2}{\pi^2 \nu^2}}
                                   \nonumber \\
                                  &-& \frac{gt}{\sqrt{\pi^3 \nu^3 {\bar n}} } \;  
                                  e^{ - \frac{1}{8} \, ( \frac{gt}{2 |\alpha|^2} )^2 } 
                                  e^{ -\frac{(gt- g{\bar t}_{\nu})^2}{2 \pi^2 \nu^2}}
                                  \sin \bigg ( \frac{gt}{2 \sqrt{ |\alpha|^2}} \bigg ) 
                                  \sin \bigg ( \frac{(gt)^2}{2 \pi \nu} - \frac{\pi}{4} \bigg ) ~ 
                        \bigg ] ~,
\eea

\noi
where
\be
  g {\bar t}_{\nu} = 2 \pi \nu \sqrt{{\bar n}} ~~.
\ee
\noi
     As shown in  \fig{tr_2_sml_FIG} the analytical form  \eq{tr_2_peaked_analytisk} with Eqs.(\ref{w_0_felles}) and (\ref{w_nu_poisson}) describes the actual from of $\Tr [ \;  \rho_{A}^2(t) \; ]$ with a very high numerical accuracy. If we make the photon-number distribution $p_n$ more peaked as compared to e.g. a Poissonian distribution we now see from \eq{tr_2_peaked_analytisk} that the approach to purity at later times $t_{rev}/2+ 2\pi k$, with $k=1,2,...,$ will be more visible. In \fig{tr_2_gaussian_FIG} we illustrate this feature for a squeezed coherent state $|\gamma \rangle =S(r)|\alpha \rangle$ with the same mean-value of photons as in \fig{tr_2_sml_FIG}.
Here we consider a squeezing operator $S(r)=\exp(r(a^2-a^{\dagger 2})/2)$ and $\alpha $ real leading to the photon-number distribution \cite{Yuen76}
\be
p_n =\frac{\tanh(r)^n}{n!2^n\cosh(r)}\exp \bigg(-|\alpha|^2(1-\tanh(r))\bigg)H_n\bigg(\frac{\alpha}{\sqrt{\sinh(2r)}}\bigg)^2~~,
\ee
where $H_n$ is a Hermite polynomial. In this case the Poisson resummed expression \eq{tr_2_peaked_analytisk} reproduces the exact answer, within the numerical accuracy of \fig{tr_2_gaussian_FIG},  by making use of the fact that the photon-number distribution is well approximated by a "squeezed" Gaussian distribution with mean ${\bar n}$ and $\sigma_n^2= {\bar n}^{0.65}$ except for times close to $t_0$. For the parameters chosen the inequality \eq{condition} actually fails for $\nu =1 $. Evaluating the integral \eq{defoff} more carefully for $\nu =1 $, which e.g. can be done analytically by making use of a Gaussian approximation for $p_n$,  one restores the agreement between the exact answer and the approximation \eq{tr_2_peaked_analytisk} with high numerical accuracy.
%

In concluding we would like to comment on the nature of pure state $|\psi \rangle_{A\otimes R}$ at $t\simeq t_0=t_{rev}/2=\pi\sqrt{{\bar n}+1}/g$. We again assume  that ${\bar n}\gg 1 $ and that the conditions
\eq{condition} are fulfilled for any probability distribution $p_n$ under consideration. As an initial state we consider a general pure entangled state, i.e.
\be
 \label{psi_entangled}
  |\psi\rangle 
               = a|+ \rangle \otimes |\gamma_+ \rangle + b|-\rangle \otimes |\gamma_- \rangle~~,
\ee
\noi
where $|a|^2+|b|^2=1$ and $|\gamma_{\pm} \rangle = \sum_{n=0}^{\infty} \sqrt{p_n^{\pm}} \, 
      e^{i \alpha_n^{\pm}} |n \rangle$.
As long as $|t-t_0 |<< 2\sqrt{\bar{n}+1}/g$ we again  make use of the fact that $gt\sqrt{n+1} = gt\sqrt{n}+ \pi/2$ \cite{Gea90} and we find that
\be
 \label{psi_entangled_t}
  |\psi(t\simeq t_0)\rangle 
               \simeq (|+ \rangle + |- \rangle e^{i\omega t_0}e^{-i\Delta\alpha}) \otimes |\gamma (t\simeq t_0) \rangle ~~,
\ee
with $\Delta\alpha = \Delta\alpha^{\pm}({\bar n})$, where $\Delta\alpha^{\pm}(n)\equiv \alpha_{n}^{\pm}-\alpha_{n-1}^{\pm}$~, and
\be
|\gamma (t) \rangle = -\sum_{n=0}^{\infty}e^{-i\omega (n+1/2)t}\bigg( ae^{i \alpha_n^{+}}\sqrt{p_n^{+}}\sin (gt\sqrt{n})+ibe^{i  \alpha_{n+1}^{-}}\sqrt{p_{n+1}^{-}}\cos (gt\sqrt{n}) ~\bigg)|n \rangle ~~.
\ee
\noi In order to obtain \eq{psi_entangled_t} we only assumed that the distributions $p_n^{\pm}$ have their main contribution at the same $n={\bar n}$ around which $\Delta\alpha^{\pm}(n)$ are assumed to be slowly varying and equal when evaluated at $n={\bar n}$. Apart from the phase factor $e^{i\omega t_0}$, \eq{psi_entangled_t} agrees with the result of Ref.\cite{Gea90}. With regard to experimental realizations it is, however, important to realize that the relative phase in \eq{psi_entangled_t} depends on both $\Delta\alpha$ and $\omega t_0$. The JC-model therefore predicts disentanglement at $t\simeq t_{rev}/2$ to the pure atomic state $|\psi \rangle_{A}= (|+ \rangle + |- \rangle e^{i\omega t_0}e^{-i\Delta\alpha})/\sqrt{2}$ fairly independent of the nature of the initial pure entangled state.

In summary, we have shown that {\it exact} coherence of the atom is in general never regained
for a JC model with a general initial pure quantum state $|\gamma \rangle$ of the radiation field. One can, however, come arbitrarily close to a disentangled pure atomic state $|\psi \rangle_{A\otimes R}$ when $|\gamma \rangle$  has a general form 
$\sum_{n=0}^{\infty} \sqrt{p_n} \, e^{i\alpha_n} |n \rangle$ provided  that the probability distribution $p_n$ is sufficiently peaked around its mean value ${\bar n} \gg 1$ and  that the phases differences $\alpha_n - \alpha_{n-1}$ do not vary significantly around this peak. Under such conditions we have then derived an approximative analytical expression for the quantity
$\Tr [ ~ \rho_{A}^2(t) ~ ]$.  Hence, the quantity 
 $\Tr [ ~ \rho_{A}^2(t) ~ ]$ exhibits revivals not only for an initial coherent 
 state of the radiation field, as one is lead to believe by the results of  Ref. \cite{Gea90},
but actually for a very general set of  pures of the radiation field. By making use of the analysis of Ref.\cite{banac90} one can, in a straightforward manner, make use of the methods above and verify that this actually remains true  independent of the initial state of the atom. We have also seen that under quite general circumstances an initial pure entangled state also leads to the same purification of the atomic state.

\vspace{0.8cm}
\begin{center}
{\bf ACKNOWLEDGMENT}
\end{center}
%

   This work was supported in part by the UK Engineering and
     Physical Science Research Council, by the European Union IHP
     Network "QUEST" and the IST Network QUBIT. P.K.R. acknowledges support by the Research Council of Norway under
   contract no. 151565/432.
   B.-S.S. wishes to thank NorFA for support and G\"{o}ran Wendin and the Department of
  Microelectronics 
and Nanoscience for hospitality.
%
%
%
%
 
%
\end{document}